\documentclass[12pt]{article}
\include{epsf}
\include{aps}
\usepackage{latexsym,epsfig}
\usepackage{graphicx}

\begin{document}

\begin{center}
{\LARGE\bf General Quasiparticle Propagator and Mass Dependence in Degenerate Color Superconductivity}
\\[1cm]
Todd Fugleberg
\\ ~~ \\
{\it Department of Physics} \\
{\it Brookhaven National Laboratory, Upton, New York 11973, U.S.A.} \\
{\it email: fugle@quark.phy.bnl.gov} 
\end{center}

\begin{center}
BNL-NT-02/12
\end{center}

\begin{abstract}
The general quasiparticle propagator in dense quark matter is derived
for equal mass quarks.  Specialized to an NJL model, this propagator
includes one new condensate, $\Delta_3$, in addition to the usual CFL
condensate, $\Delta_1$.  The gap equation is solved in two NJL models
and the dependence on the quark mass of the condensates and
the gap is presented.  Analytic approximations for the condensates are
obtained and compared to exact numerical solutions.  The results are
shown to differ from those obtained by neglecting $\Delta_3$,
especially for smaller values of $\Delta_1$.  The two different NJL
models presented are also shown to give different results when
$\Delta_3$ is not neglected. The methods used in this paper can be
generalized to the physical case where only the strange quark is
significantly massive.
\end{abstract}

\renewcommand{\thesection}{\Roman{section}}

\section{Introduction}

The physics of strongly interacting matter at high densities and low
temperatures has been the subject of much research in recent years.
It has long been known\cite{Barrois,BailinLove} that at sufficiently
high densities a system of quarks should form a condensate of Cooper
pairs which breaks the $SU_C(3)$ symmetry and becomes a color
superconductor.  The formation of the diquark condensates leads to
gaps in the quasi-particle spectrum.  The authors of
\cite{Barrois,BailinLove} estimated that the gaps were of the order of
$\Delta\sim 10^{-3}\mu$ where $\mu$ is the quark chemical
potential. More recently it was shown at realistic values of $\mu$ in an
instanton induced NJL\footnote{The designation NJL (Nambu-Jona-Lasinio)
model is used in this paper to describe models where the quark
interaction is taken to be a four-fermion interaction at a point.} 
model that gaps of the order of 10-100 MeV could be
obtained\cite{ARW_RSSV}.  This stimulated a great deal of
research\footnote{For extensive references see the review articles
\cite{Review1,Review2}.} in the ensuing years and has resulted in a
proliferation of predicted superconducting ground states.  These
states may be realized in the cores of neutron or quark stars and lead to
observable effects\cite{CCSCCompact,Carter_Blinking,Prakash,Pons_etal}.

It is widely accepted\cite{CFL,CFL_accepted} that for three colors and
flavors the color superconducting ground state at asymptotic
densities is the Color Flavor Locked (CFL) state\cite{CFL}:
\begin{eqnarray}
\langle q_\alpha^i C \gamma^5 q_\beta^j  \rangle&=&  \Delta_{\bar{3}\bar{3}} 
(\delta^i_\alpha \delta^j_\beta-\delta^i_\beta\, \delta^j_\alpha)
+\Delta_{66}
(\delta^i_\alpha \delta^j_\beta+\delta^i_\beta\, \delta^j_\alpha)  
\label{CFL}\\
&\approx& \Delta_{\bar{3}\bar{3}} \left(
\sum_{A=2,5,7} \lambda^A_{\alpha\beta} \lambda^A_{ij} \right),
\nonumber
\end{eqnarray}
where the Greek indices are color indices, the Latin indices are
flavor indices, the $\bar{3}$ and $6$ subscripts refer to anti-triplet
or sextet configurations in color and flavor spaces respectively
and $\lambda^A$ are the Gell-Mann matrices.
At lower densities 
($\mu\sim m_s^2/4 \Delta$), it is likely that the ground
state is a superconducting state involving the condensation
of Cooper pairs in the $u-d$ sector only (2SC):
\begin{equation}
\langle q_\alpha^i C \gamma^5 q_\beta^j  \rangle= \Delta_{2SC} \,
 \epsilon_{\alpha\beta 3}\,\epsilon^{ij3},
\label{2SC}
\end{equation}
Finally at still lower densities the favored ground state will be
ordinary hadronic matter.  Two new phases have recently been
predicted: Crystalline Color Superconductivity\cite{Crystalline} 
and CFL with meson condensation\cite{Stress,Kaplan_Reddy}.
These predictions, while not necessarily at odds with one another,
indicate that the transition region between the CFL state and hadronic
matter is not completely understood.

A standard way of studying color superconductivity is to solve the gap equation
in the Nambu Gorkov formalism which is reviewed in Appendix A.
The inclusion of a strange quark mass in the gap equation
introduces two sets of complications: 1) massive quarks
means that there are 4 new allowed Dirac structures for the
condensates, and coupling between the condensates; 2) the fact that the
strange quark is different from the other quarks means that
condensates involving the strange quark should be different from those
with zero strangeness.

In order to understand the implications of these two complications it
is useful to separate them and understand them individually
before tackling the full problem.   In a previous paper by this
author\cite{Fugleberg} the second problem of non-degenerate quarks was 
studied in the case where all quarks are massless but the
strange quark is given a different chemical potential than the
other two quarks.  This paper concentrates on the first complication
by considering the problem of equal mass quarks. 

Color superconductivity with massive quarks has been studied before.
The papers \cite{Unlocking_ABR}-\cite{Mass_induced}
have dealt with inclusion of quark mass or strange quark mass in NJL models.  
The papers \cite{Stress,Kaplan_Reddy,MassTerms} have dealt with inclusion of quark
mass or strange quark mass in effective Lagrangian models.  Finally
\cite{Opening,Harada} have dealt with this problem in the perturbative
approach.  All of these important papers have dealt with the problem
at some level of approximation.  The new results in this paper are
solutions for all allowed diquark condensates applicable over a wide
range of quark masses and coupling constants.  The quasiparticle
propagator derived herein is the general form for the case of equal
mass quarks.  It is shown that for the NJL model there is one and only
one additional diquark condensate, $\Delta_3$, and it is included in
the calculations.  The approximate analytic solutions presented in
this paper make numerical solution of gap equations
for different parameter values unnecessary.  Finally two different NJL
models are analyzed in exactly the same framework so that they can be
compared and contrasted.

The authors of \cite{Unlocking_ABR} did discuss the fact that
$\Delta_3$, which they refer to as $\kappa$, must be non-zero but
since it is small they neglect it in their calculations.  The authors
of \cite{Buballa_Hosek_Oertel} include a similar diquark condensate in
the 2SC phase, but their analysis does not indicate that no other
diquark condensates would arise. In \cite{Buballa_Hosek_Oertel} it was
shown that $\Delta_3$ can be large in an NJL model using a Lorentz
non-invariant interaction and therefore have a significant effect on
the gap.  In this paper it is shown that, even for small values of
$\Delta_3$, neglecting this condensate in the calculation can have a
non-trivial effect on the gap.

The results of this paper are an approximation to the $N_f=3$ CFL case
as the $\Delta_{66}$ gaps are neglected to simplify the presentation.
The results are exact for $N_f=2$ and for the 2SC phase
with $N_f=3$.  Therefore there is only one color-flavor
structure and we drop the subscripts ``$\bar{3}\bar{3}$'' and ``2SC''
as they are unnecessary.

The general quasiparticle propagator is presented for the case of
equal mass quarks.  This propagator is then specialized to the NJL
model where it is shown that there is only one new diquark condensate,
$\Delta_3$, in addition to the CFL condensate, $\Delta_1$.

The gap equation was solved in two different NJL models.  The first
model used is the simplest possible NJL model equivalent to scalar
exchange.  The second model used is an NJL model with the color
structure of single gluon exchange.  The second model is perhaps more
physically motivated and is the model usually used in NJL analyses of
color superconductivity.  However, there is no reason to exclude
scalar interaction terms and restrict the analysis to interactions
motivated by single gluon exchange as a general low energy Lagrangian
could include both.  This being said it is interesting to see how much
of an effect the specific NJL model has on the results.

The gap equation was solved in the scalar NJL model and results are
presented for the general set of condensates as a function of the
quark mass.  Approximate analytic solutions for the condensates are
obtained and shown to agree well with exact numerical solutions.  The
results are compared with results obtained neglecting the effect of
the new condensate $\Delta_3$.  It is shown that the inclusion of
$\Delta_3$ alters the results especially for lower values of
$\Delta_1$.  The gap in the quasiparticle spectrum is shown to be a
function of both $\Delta_1$ and $\Delta_3$ and inclusion of $\Delta_3$
is even more relevant.

The same analysis was performed in an NJL model with the color
flavor structure of single gluon exchange.  The results in this model
are also altered by the inclusion of $\Delta_3$ especially for lower
values of $\Delta_1$ but they have the opposite effect as in the
scalar NJL model.

These results are relevant to the analysis of
\cite{Huang_propagator} where the gap equation is solved neglecting
$\Delta_3$.  Their results are shown to be a reasonable approximation
for the parameters they chose, but can be generalized using the
techniques of this paper.  The effects of chiral symmetry breaking
have not been taken into account in this paper as in
\cite{Huang_propagator,Buballa_Hosek_Oertel} but such
 an extension would not be difficult.

These results illustrate the effect of using the general form of the
quasiparticle propagator in the solution of the gap equations for
massive but degenerate quarks.  The results are most significant for
smaller values of the CFL condensate and they depend strongly on the
specific NJL model used.

The methods used in this paper are a significant new result on their
own.  They can be generalized to determine the general quasiparticle
propagator and the general set of gap equations for the physical case
where the up and down quarks are essentially massless, but the strange
quark is massive.  As well, this analysis could be repeated using
perturbation theory in order to compare with a model that is exact at
higher densities and energies.

The outline of this paper is as follows. In section II a basis for the
Dirac structure of the condensates is presented which facilitates the
rest of the analysis.  In section III the general quasiparticle
propagator is derived and then specialized to the NJL model.  In
section IV the gap equations are solved in the scalar NJL model.
In section V the gap equations are solved in an NJL model which
has the color structure of single gluon exchange.  Section VI concludes
with a summary of the results.  Appendix A contains a brief review
of the Nambu-Gorkov formalism.  Appendix B gives the decomposition
from \cite{BailinLove} of the gap matrix in terms of the products of Dirac matrices.

\section{Basis for the Condensates}

The CFL condensate in (\ref{CFL}) has the Dirac structure $\gamma_5$
and is the only spin zero condensate for the massless case in an NJL
model.  The most general set of spin zero condensates is given in
Appendix B and includes eight different possible structures.  In an
NJL model the structures involving $\gamma\cdot\hat{k}$ automatically
vanish leaving four possible Dirac structures:
$\gamma_5,\gamma_5\gamma_0,\gamma_0,I$.  Solving the gap equations
give the result that there is only one new condensate, $\Delta_3$
with the Dirac structure: $\gamma_5\gamma_0$.

The analysis could be carried out in a basis of Dirac structures, but
the equations would be very complicated.  A crucial step in finding
the general quasiparticle propagator and gap equations is choosing a
basis which simplifies the analysis enough to make it tractable.

In the massless case an arbitrary condensate matrix can be decomposed using
a basis of orthogonal projection operators\cite{RD_superfluid}:
\begin{equation}
P^e_h(\vec{k})=\Lambda^e(\vec{k}) P_h(\vec{k}), ~~~~ e,h=\pm 1,
\end{equation}
which are products of positive and negative energy projectors:
\begin{equation}
 \Lambda^e(\vec{k})=
\frac{1+e\,\gamma_0\vec{\gamma} \cdot {\vec k}}{2} ~~~~ e=\pm 1,
\end{equation}
and positive and negative helicity projectors:
\begin{equation}
P_h(\vec{k})=\frac{1+e\,\gamma_5\gamma_0\vec{\gamma} \cdot {\vec k}}{2}  ~~~~ h=\pm 1.
\end{equation}
In this basis the condensate matrix is:
\begin{equation}
\Delta= \sum_{e,h,\pm1} \Delta^e_h \; P^e_h.
\end{equation}
Similarly all objects, such as the bare quark and the quasiparticle propagator can
be written in this basis.  The basis of orthogonal projection matrices is extremely useful
since it reduces products of these objects to the simplest possible form.

In the case of massive quarks things are more complicated.
The energy projector can be generalized to\cite{RD_superfluid}:
\begin{equation}
\Lambda^e(\vec{k})=
\frac{1+e\,\left(\beta \gamma_0\vec{\gamma} \cdot {\vec k} + \alpha \gamma_0 \right)}{2}  ~~~~ e=\pm 1,
\end{equation}
where $\beta\equiv |\vec{k}|/E_k$ and $\alpha \equiv m/E_k$.
A set of analogous operators in the massive case given in \cite{RD_superfluid}
is:
 \begin{equation}
P^e_{c\,h}(\vec{k})= P_c \, \Lambda^e(\vec{k}) P_h(\vec{k}), ~~~~ e,h=\pm 1 ~~~ c=r,l,
\end{equation}
where:
\begin{equation}
P_c=\left(\frac{1+c\,\gamma_5}{2} \right) ~~~~ c=r,l,
\end{equation}
is the chirality projector which projects onto right and left handed
spinors.  The definition $r=+$ and $l=-$ in analogy with the other
projectors will be used in what follows.
These operators are really quasiprojectors with the general
product rule:
\begin{equation}
P^e_{ch}(k) P^{e'}_{c'h'}(k)=\delta_{ee'}\delta_{cc'}\delta_{hh'} P^e_{ch}(k)
-\frac{1}{2} e\,c\,e'c' (1- e' c\,h\,\beta) \, \delta_{hh'}\,P^{e'}_{ch}(k).
\end{equation}
They are not projectors because chirality and energy projectors do not commute:
\begin{equation}
\left[ P_c, \Lambda^e\right]= -e\,c\; \alpha\; \gamma_0\gamma_5.
\end{equation}
This basis is complete and the general gap matrix can be written\cite{RD_superfluid}:
\begin{equation}
\Delta= \sum_{e,c,h,\pm1} \Delta^e_{c\,h} \; P^e_{c\,h}.
\end{equation}
This basis still involves many complications.

The calculation is simplified by defining a new basis of true
projectors and nilpotent operators:
\begin{equation}
P^e_h(k)= \sum_{c=-1,1} P^e_{ch}(k)= P_h(k) \: \Lambda^e(k),
\end{equation}
\begin{equation}
Q^e_h(k)= \sum_{c=-1,1} (e\,c\,h-\beta) P^e_{ch}(k)= (e\,h\,\gamma^5-\beta)P^e_h(k),
\end{equation}
that satisfy the following relations:
\begin{eqnarray}
P^e_h(k) P^{e'}_{h'}(k)&=&\delta_{ee'}\delta_{hh'} P^e_h(k), \\
Q^e_h(k) P^{e'}_{h'}(k)&=&\delta_{ee'}\delta_{hh'} Q^e_h(k), \\
P^e_h(k) Q^{e'}_{h'}(k)&=&\delta_{-ee'}\delta_{hh'} Q^{-e}_h(k),\\
Q^e_h(k) Q^{e'}_{h'}(k)&=&-\alpha^2\delta_{-ee'}\delta_{hh'} P^{-e}_h.
\end{eqnarray}
The $P^e_h(k)$ are orthogonal projectors and the $Q^e_h(k)$ are
nilpotent even though products of $P^e_h(k)$ and $Q^e_h(k)$ or
$Q^e_h(k)$ and $Q^{e'}_{h'}$ mix the operators. $P^e_h(k)$ is an
hermitian operator and $(Q^e_h(k))^\dagger=-Q^{-e}_h(k)$.  In the
chiral limit ($\alpha\rightarrow 0$, $\beta\rightarrow 1$) $P^e_h(k)$
are identical to the massless projectors and $Q^e_h(k)\rightarrow 0$.
Notice that the operators with different helicities are
completely decoupled.

The multiplication table for each helicity is:
\begin{eqnarray}
~~~~~~~~~~~P^+_h ~~~~~~ Q^+_h ~~~~~ P^-_h ~~~~~~~~ Q^-_h \nonumber
\end{eqnarray}
\vspace{-0.36in}
\begin{eqnarray}
\begin{array}{c}
P^+_h \\ 
Q^+_h \\
P^-_h \\
Q^-_h
\end{array} 
&
\left(
\begin{array}{lccr}
P^+_h & 0               & 0     & Q^-_h           \\
Q^+_h & 0               & 0     & -\alpha^2 P^-_h \\
0     & Q^+_+           & P^-_h & 0               \\
0     & -\alpha^2 P^+_h & Q^-_h & 0 
\end{array} \right).
\label{Multiplication_Table}
\end{eqnarray}

This set of eight operators is the set of operators on the space of
solutions of the Dirac equation which gives the most sparse
multiplication table.  One would prefer a basis of only orthogonal
projection operators as in the massless case, but such a set does not
exist.  Dirac spinors are 4 component complex vectors and as such live
in ${\cal R}^8$.  One can definitely find eight orthogonal projectors
in this space, but Dirac spinors must be solutions of the Dirac
equation which places constraints on the components and reduces the
dimensionality of the space.  Therefore, you cannot have 8 orthogonal
projectors onto the space of solutions of the Dirac equation.  On the
other hand you need 8 operators to form a complete set with which to
construct all 8 independent Dirac structures.  In the massless case,
the left and right handed spinors decoupled so that each was a two
component complex spinor living in ${\cal R}^4$ with two constraints
from the Dirac equation.  It is therefore possible to find a set of 4
orthogonal projectors on this space.

The operators given above have the following representation as $2\times 2$ matrices:
\begin{eqnarray}
P^+_h= \left(
\begin{array}{lr}
1 & 0 \\
0 & 0
\end{array}
\right), & 
P^-_h= \left(
\begin{array}{lr}
0 & 0 \\
0 & 1
\end{array}
\right),
\end{eqnarray}
\begin{eqnarray}
Q^+_h= \left(
\begin{array}{lr}
0 & 0 \\
-\alpha & 0
\end{array}
\right), & 
Q^-_h= \left(
\begin{array}{lr}
0 & \alpha \\
0 & 0
\end{array}
\right).
\end{eqnarray}
This supports the reasoning above that there are only
4 independent projectors since the operators for each handedness operate
on a separate two dimensional space.  This representation will be useful in what follows.

This set of operators is complete as shown by the following
relations:
\begin{eqnarray}
I&=&\sum_{e,h=-1,1} P^e_h(k), \\
\gamma^5 \gamma^0 \gamma \cdot {\hat k}&=&\sum_{e,h=-1,1} h \:P^e_h(k), \\
\alpha\, \gamma \cdot {\hat k}&=&-\sum_{e,h=-1,1} Q^e_h(k), \\
\alpha\, \gamma^5\gamma^0&=&\sum_{e,h=-1,1} h \:Q^e_h(k), \\
\gamma^0 \gamma \cdot {\hat k}&=&\sum_{e,h=-1,1} e \,(Q^e_h(k) + \beta\,P^e_h(k)),\\
\gamma^5&=&\sum_{e,h=-1,1} e\,h \,(Q^e_h(k) + \beta\,P^e_h(k)),\\
\alpha \,\gamma^0&=&\sum_{e,h=-1,1} e \,(\alpha^2 P^e_h(k) - \beta\,Q^e_h(k)),\\
\alpha \,\gamma^5 \gamma \cdot {\hat k}&=&-\sum_{e,h=-1,1} e \,h\,(\alpha^2 P^e_h(k) - \beta\,Q^e_h(k)).
\end{eqnarray}
It should be noted that these equations are still valid in the chiral
limit (albeit half of them are trivial).

The gap matrix is decomposed in terms of these operators as:
\begin{eqnarray}
\Delta=\sum_{e,h=-1,1} \left(\beta\;\xi^e_h+\alpha^2\;\psi^e_h\right) P^e_h(k) + \left(\xi^e_h-\beta\;\psi^e_h\right) Q^e_h(k), \\
\Delta^\dagger=\sum_{e,h=-1,1} \left(\beta\;\xi^e_h+\alpha^2\;\psi^e_h\right) P^e_h(k) - \left(\xi^e_h-\beta\;\psi^e_h\right) Q^{-e}_h(k).
\end{eqnarray}

All objects in this analysis can be represented
using this basis.  The sparse multiplication table simplifies 
much of the analysis in the rest of this paper.  In the next section 
the general quasiparticle propagator is derived using this basis.

\section{General Quasiparticle Propagator}

The quasiparticle propagator is determined by the 
equation\cite{RD_superfluid}\footnote{see Appendix A for a short review of the
Nambu-Gorkov formalism.}
\begin{equation}
G^\pm\equiv\left\{ [G^\pm_0]^{-1}-\Delta^\mp G^\mp_0 \Delta^\pm\right\}^{-1}.
\label{quasiparticle_propagator}
\end{equation}
where $\Delta^+\equiv\Delta$, $\Delta^-\equiv \gamma_0 \Delta^+ \gamma_0$
and $G^-_0$ is the bare charge-conjugate particle propagator:
\begin{eqnarray}
G^-_0(k)&=&
\left(\gamma^\nu k_\nu-\mu \gamma^0- M \right)^{-1}=
\frac{\gamma^\nu k_\nu-\mu \gamma^0+ M}{(k_0-\mu)^2-E_k^2} \\
&=& \gamma^0
\frac{\left((k_0-\mu)-E_k \beta\gamma^0\gamma \cdot {\hat k}+E_k\alpha\gamma^0  \right)}{(k_0-\mu)^2-E_k^2} \nonumber \\
&=&\gamma^0\frac{(k_0-\mu)+E_k\sum_{e,h=-1,1} e \,\left[(\alpha^2-\beta^2) P^e_h(k) - 2\beta\,Q^e_h(k)\right]}{(k_0-\mu)^2-E_k^2}. \nonumber
\end{eqnarray}
Defining:
\begin{eqnarray}
\Omega&=& \left[G^+(k)\left((k_0-\mu)\gamma^0-\gamma \cdot {\vec k}+ M \right)^{-1} 
\right]^{-1}\nonumber\\ 
&=& \sum_{e,h=-1,1}\left(A^e_h P^e_h(k) + B^e_h Q^e_h(k) \right) 
= \sum_{h=-1,1}\Omega_h, 
\label{Omega}
\end{eqnarray}
the coefficients are:
\begin{eqnarray}
A^{e}_{h}&=& (k_0)^2- \epsilon^{e}(\xi^{e}_{h},\psi^{e}_{h})^2 - \frac{2\, e\,\alpha_k^2\, E_k}{(k_0-\mu)- e E_k}
(\psi^e_h)^2,  \\
B^{e}_{h}&=&\xi^{-e}_{h} \psi^{e}_{h} - \xi^{e}_{h} \psi^{-e}_{h}
+\frac{2\, e\, E_k}{(k_0-\mu)+ e E_k}\xi^{e}_{h}\, \psi^{-e}_{h}, \nonumber
\end{eqnarray}
using the notation:
\begin{equation}
\epsilon^\pm(\xi,\psi)= \left[ (E_k\mp\mu)^2 + 
\xi^2 + \alpha_k^2 \,\psi^2 
\right]^{1/2}.
\label{old_epsilon}
\end{equation}
Using the $2\times 2$ matrix representation of $P^e_h$ and $Q^e_h$ given above:
\begin{equation}
\Omega^{-1}= \sum_{e,h=-1,1}\frac{1}{\det \Omega_h}\left(A^{-e}_h P^e_h(k) - B^e_h Q^e_h(k) \right),
\end{equation}
where:
\begin{eqnarray}
\det \Omega_h&=&k_0^4 -2 k_0^2\left(E_k^2+ \mu^2 + (\xi^+_h)^2 + (\xi^-_h)^2 + \alpha^2(\psi^+_h)^2 + \alpha^2(\psi^-_h)^2\right) \\
&+& 2 E_k k_0\left((\psi^+_h)^2 - (\psi^-_h)^2\right)\nonumber \\
&+&\left[(E_k-\mu)^2 (E_k+\mu)^2 +  (E_k+\mu)^2 (\xi^+_h)^2+(E_k-\mu)^2 (\xi^-_h)^2 \right.\nonumber \\
&-& \left.\alpha^2 (E_k-\mu) (E_k+\mu) \left((\psi^+_h)^2 - (\psi^-_h)^2 \right) +\left(\xi^+_h \xi^-_h+ \alpha^2 \psi^+_h\psi^-_h\right)^2\right], \nonumber
\end{eqnarray}
and the definition:
\begin{equation}
E_k= \sqrt{|\vec{k}|^2+m^2},
\end{equation}
has been used.  The poles of the quasiparticle propagator are given by the roots of this quartic equation in $k_0$.  The roots have well known
analytic forms that are nonetheless quite complicated and are
not necessary for the purposes of this paper.

From (\ref{Omega}) is it clear that:
\begin{equation}
G^+(k)=\Omega^{-1} \left((k_0-\mu)\gamma^0-\gamma \cdot {\vec k}+ M \right),
\end{equation}
and therefore the general quasiparticle propagator is:
\begin{eqnarray}
G^+(k)&=& 
\label{completely_general_propagator}\\
&& \!\!\!\!\!\!\!\!\!\!\!\!\!\!\!\!\!\!\!\!\!\!\!\!\!\!\!\!\!\!\!\sum_{e,h=-1,1}\frac{\alpha (e\;k_0-e\;\mu + E_k))(k_0^2-(\omega_1)^e_{h})-\alpha(e\; k_0-e\;\mu-E_k)\left( \beta\;\upsilon^e_h\right) \psi^{-e}_h}
{\det \Omega_h} P^e_h \nonumber \\
&&\!\!\!\!\!\!\!\!\!\!\!\!\!\!\!\!\!\!\!\!\!\!\!\!\!\!\!\!\!-\sum_{e,h=-1,1}\frac{\beta/\alpha(e\; k_0-e\;\mu-E_k)(k_0^2-(\omega_2)^e_{h})+\alpha(e\;k_0-e\;\mu- E_k)\eta^e_h \psi^{-e}_h}
{\det \Omega_h} Q^e_h. \nonumber 
\end{eqnarray}
where:
\begin{eqnarray}
(\omega_1)^e_{h}&=&(E_k+ e\; \mu)^2 +\xi^{-e}_h \chi^{-e}_h, \\
(\omega_2)^e_{h}&=&(E_k+ e\; \mu)^2 + \frac{1}{\beta}\;\xi^{-e}_h \phi^{-e}_h,
\end{eqnarray}
and:
\begin{eqnarray}
\phi^e_h&=& \beta\;\xi^e_h+\alpha^2\;\psi^e_h, \\
\chi^e_h&=&\xi^e_h-\beta\;\psi^e_h, \\
\upsilon^e_h&=& \beta\;\xi^e_h+\alpha^2\;\psi^{-e}_h, \\
\eta^e_h&=&\xi^e_h-\beta\;\psi^{-e}_h.
\end{eqnarray}
This propagator could be diagonalized but the
diagonalization depends on $k_0$ and $E_k$ and is quite complicated.
It is not necessary to diagonalize the complete propagator
for the purposes of this paper.

Assuming that only the $\gamma_5$ condensate contributes, this corresponds to the
relations:
\begin{eqnarray}
\xi^e_h\equiv\chi^e_h \equiv \beta \phi^e_h = e\;h\; \xi^+_+, & ~~ \psi^e_h\equiv 0, & 
\upsilon^e_h\equiv\beta\xi^e_h, ~~~ \eta^e_h\equiv\xi^e_h,
\end{eqnarray}
and it can be shown that:
\begin{eqnarray}
G^+&=& \sum_{e,h=-1,1}  \frac{k_0+e (E_k-e\;\mu)}{k_0^2- (E_k-e\;\mu)^2-(\xi^+_+)^2} P^e_h\; \gamma_0 \\
&=& \sum_{e=-1,1} \frac{k_0+e (E_k-e\;\mu)}{k_0^2- (E_k-e\;\mu)^2-(\xi^+_+)^2} \Lambda^e \;\gamma_0, \nonumber
\end{eqnarray}
which agrees with the quark propagator derived in
\cite{Huang_propagator} using this ansatz.  While this ansatz is a
good first approximation, it will be shown that the gap equation will
not close under this ansatz and the general quark propagator is
the more complex form shown in (\ref{completely_general_propagator}).

The analysis up until this point generalizes to perturbation theory.
In what follows the analysis will be restricted to NJL models where
things simplify considerably.
With foresight define:
\begin{eqnarray}
 \psi^{e}_{h}&=&h \; \frac{\beta}{\alpha} \; \Delta_3, \\
\xi^{e}_{h}&=& h\;(e \Delta_1-\alpha \Delta_3),
\end{eqnarray}
where $\Delta_1$ is the usual CFL condensate and $\Delta_3$ is a new condensate
and the gap matrix is now given by:
\begin{equation}
\Delta=\Delta_1 \gamma_5 + \Delta_3 \gamma_5\gamma_0.
\end{equation}
In this case:
\begin{equation}
\det \Omega_+=\det \Omega_-=\left(k_0^2-(\varepsilon^+)^2 \right)\left(k_0^2-(\varepsilon^-)^2\right), 
\end{equation}
where the poles of the quasiparticle propagator are given by:
\begin{equation}
(\varepsilon^\pm)^2= E_k^2 + \mu^2 +\Delta_1^2+ \Delta_3^2 
\pm \sqrt{\left( 2 E_k \mu +2 \alpha\Delta_1 \Delta_3 \right)^2
+  4\beta^2 \Delta_3^2 \left(  E_k^2 + \Delta_1^2 \right)}\;.
\label{poles}
\end{equation}
Note that $\varepsilon^\pm$ are distinct from the $\epsilon^\pm$
defined in (\ref{old_epsilon}) and are used exclusively in what
follows.  These dispersion relations can be shown to agree with
those stated in \cite{Buballa_Hosek_Oertel}.

The poles of the propagator give the dispersion relations for quasiparticles,
(-$\varepsilon^-$), quasi-antiparticles (-$\varepsilon^+$),
quasiparticle holes ($\varepsilon^-$) and quasi-antiparticles holes
($\varepsilon^+$).  The physical manifestation of the condensates is a gap in the
quasiparticle spectrum between the maximum of the quasiparticle branch
and the minimum of the quasiparticle hole branch:
\begin{equation}
\varphi=(\varepsilon^-|_{min})-(-\varepsilon^-|_{max})=2 \varepsilon^-|_{min}.
\end{equation}
This means that exciting a quasiparticle requires a minimum energy, $\varphi$.

In the massless limit $\Delta_3=0$,
$\alpha=0$, $E_k=|\vec{k}|$
and therefore the $\varepsilon^\pm$ reduce to $\epsilon^\mp$ defined in
\cite{RD_superfluid}.  In this case the gap is simply  
$\varphi=2\Delta_1$ at $k=\mu$.

In the massive case, $\Delta_3$ alters the dispersion relations. The minimum
of the quasiparticle hole spectrum occurs at:
\begin{equation}
E_k=\sqrt{\mu^2 +\Delta_1^2+\Delta_3^2 - \frac{2 m\mu \Delta_1+ (\Delta_1^2  - m^2) \Delta_3}{\Delta_3^2+\mu^2}}\;,
\end{equation} 
and leads to a gap:
\begin{equation}
\varphi=2\;\frac{\mu\Delta_1-m\Delta_3}{\sqrt{\mu^2+\Delta_3^2}}\;.
\label{gap}
\end{equation}
For $\Delta_3=0$ the minimum occurs at $E_k=\mu$ and leads to
a gap of $\varphi=2\Delta_1$. When $\Delta_3=0$ is taken into account
the minimum is shifted to a slightly lower value and the gap increases
or decreases depending on the sign of $\Delta_3$. Further discussion of
the dispersion relations and the gap will be given in subsequent sections
in terms of specific solutions.

In the NJL model the full propagator is;
\begin{eqnarray}
G^+(k)&=&\label{NJL_full_propagator}\\
&&\!\!\!\!\!\!\!\!\!\!\!\!\!\!\!\!\!\!\!\!\!\!\!\!\!\!\!\!\!\!\!\sum_{e,h=-1,1}\frac{\alpha\; e\;(k_0-\mu + e\;E_k))(k_0^2-(\omega_1)^e_{+})-e( k_0-\mu-e\;E_k) \beta^3 \Delta_1 \Delta_3}
{\left(k_0^2-(\varepsilon^+)^2 \right)\left(k_0^2-(\varepsilon^-)^2\right)} P^e_h \nonumber \\
&&\!\!\!\!\!\!\!\!\!\!\!\!\!\!\!\!\!\!\!\!\!\!\!\!\!\!\!\!\!\!\!\!
-\frac{\beta/\alpha(e\; k_0-e\;\mu-E_k)(k_0^2-(\omega_2)^e_{+})+(k_0-\;\mu- e\;E_k)(\Delta_1-e\Delta_3/\alpha) \Delta_3}
{\left(k_0^2-(\varepsilon^+)^2 \right)\left(k_0^2-(\varepsilon^-)^2\right)} Q^e_h, \nonumber 
\end{eqnarray}
where:
\begin{eqnarray}
(\omega_1)^e_{h}&=&(E_k+ e\; \mu)^2 + (\Delta_1+e\;\alpha \Delta_3)(\Delta_1+e\; \Delta_3/\alpha), \\
(\omega_2)^e_{h}&=&(E_k+ e\; \mu)^2 + \;(\Delta_1+e\;\alpha \Delta_3) \Delta_1.
\end{eqnarray}

The quasiparticle propagator can now be used in the
mean field gap equation\cite{RD_superfluid}(\ref{GapEquation1_Appendix}):
\begin{equation}
\Delta(k)= - i g^2 \int \frac{d^4q}{(2 \pi)^4} \sum_{A,B=1..N_c^2-1} {\bar \Gamma}^A_\mu D^{\mu\nu}_{AB}(k-q)
 G^-_0(q) \Delta(q) G^+(q) \Gamma^B_\nu,
\end{equation}
where $D^{\mu\nu}_{AB}(k-q)$ is the gluon propagator, 
$\Gamma^B_\nu$ is the interaction vertex, and(see Appendix A):
\begin{equation}
{\bar \Gamma}= C  \Gamma^T  C^{-1}.
\end{equation}

In the following sections results are presented for two different
types of four fermion interactions.  The next section presents results
for the scalar NJL model.  Section V presents results for an NJL model
which has the color structure of single gluon exchange.  These models
give identical results if $\Delta_3$ is neglected but differ quite
significantly if it is included.

One of the main results of this section is the general quasiparticle
propagator (\ref{completely_general_propagator}) which could be used
in a perturbation theory analysis of the gap equation.  The second
main result of this section is the quasiparticle propagator for an NJL
model (\ref{NJL_full_propagator}) which is used in the rest of this
paper.  The poles of the quasiparticle propagator (\ref{poles}) are
particularly important for this analysis.  Finally the definition of
the gap in terms of the condensates (\ref{gap}) is another result
which will be important in the rest of this paper.

\section{Scalar NJL Model}

The simplest possible four fermion interaction, motivated by the effective Lagrangian
approach, is to take the interaction vertex to be $\Gamma^A_\mu=i$ and:
\begin{equation}
g^2 D^{\mu\nu}_{AB} \rightarrow \frac{2\pi^2}{N_c^2-1} G \; \delta^{\mu\nu} \delta_{AB},
\end{equation}
giving the matrix gap equation:
\begin{equation}
\Delta(k)= 8\; i\;\pi^2\; G \int \frac{d^4q}{(2 \pi)^4} G^-_0(q) \Delta(q) G^+(q).
\label{GapEquation2}
\end{equation}

Acting on both sides of this equation with the operators $\gamma_5$,
$\gamma_5\gamma_0$ and tracing over the spinor indices one can obtain
the gap equations for $\Delta_1$ and $\Delta_3$ respectively.  The
color-flavor structure of the gap matrix can almost be ignored because
the propagators do not have any color flavor structure and therefore
the same factor will occur on both sides of the equation.  If
one is working in the degenerate three flavor case there is a slight
complication to even the zero mass gap equation because of the existence of
the $\Delta_{66}$ condensate.  This complication is ignored
in this paper so this analysis is an approximation in the
three flavor case.

In the two flavor case there is only one color-flavor structure to be concerned about
and this analysis is exact.  This applies even to the case of the 2SC color
superconducting phase in the physical 3 flavor case as condensates involving
the third color and flavor will simply decouple from the condensates of interest.

The coupled gap equations are:
\begin{eqnarray}
\Delta_1&=&8\; i\; G\pi^2 \int \frac{d^4q}{(2 \pi)^4}\frac{ 2 E_q\mu \alpha \Delta_3 +\Delta_1 (q_0^2-E_q^2-\Delta_1^2+\Delta_3^2-\mu^2)}{\left(q_0^2-(\varepsilon^+)^2\right)\left(q_0^2-(\varepsilon^-)^2\right)}, \\
\Delta_3&=&-8\; i\; G\pi^2 \int \frac{d^4q}{(2 \pi)^4} \\
&& \!\!\!\!\!\!\!\!\!\frac{ 2 E_q\mu \alpha \Delta_1 +\Delta_3 (q_0^2+\Delta_1^2-\mu^2) +\Delta_3 (\beta^2-\alpha^2) (E_q^2-\Delta_3^2)}{\left(q_0^2-(\varepsilon^+)^2\right)\left(q_0^2-(\varepsilon^-)^2\right)}, \nonumber
\end{eqnarray}
where terms linear in $q_0$ have been dropped since they will cancel
out on integration over $q_0$ from $-\infty$ to $\infty$.  The dependence of the
condensate on momentum has been dropped since the right hand side of the gap equations
are independent of $\vec{k}$.

Acting on (\ref{GapEquation2}) with the operators $\gamma_0$ and the identity and taking the trace leads to vanishing of the right hand side which is consistent with the assumption that $\Delta_4$ and $\Delta_8$ are zero.  Acting with any of the other operators involving $\gamma\cdot\vec{k}$
will lead to a term involving $\hat{q}\cdot\hat{k}$ which will vanish by symmetry
under the angular integration.

Evaluation of these equations can be facilitated by the analytic continuation
$q_0 \rightarrow -i q_4$.  The $q_4$ integration is then done by contour integration
closing the contour in the upper half plane and picking up the poles at $i \varepsilon^+$
and $i \varepsilon^-$.  The angular integrals can be done trivially giving:
\begin{equation}
\Delta_1= G \int dq \; q^2\;\left(\frac{\Delta_1}{\varepsilon^+} + \frac{\Delta_1}{\varepsilon^-}+\frac{4(m\; \mu+\Delta_1\Delta_3)\Delta_3}
{((\varepsilon^+)^2-(\varepsilon^-)^2)\varepsilon^+} -\frac{4(m\; \mu+\Delta_1\Delta_3)\Delta_3}
{((\varepsilon^+)^2-(\varepsilon^-)^2)\varepsilon^-} \right),
\label{Delta1_gap_equation}
\end{equation}
\begin{eqnarray}
\Delta_3&=& G \int dq \; q^2\;\left(-\frac{\Delta_3}{\varepsilon^+} - \frac{\Delta_3}{\varepsilon^-}-\frac{4\; m\; \mu\;\Delta_1}
{((\varepsilon^+)^2-(\varepsilon^-)^2)\varepsilon^+} +\frac{4\; m\; \mu\;\Delta_1}
{((\varepsilon^+)^2-(\varepsilon^-)^2)\varepsilon^-} \right.  \nonumber \\
&-&\left. \frac{4 \left(q^2+\Delta_1^2 \right)\Delta_3}{((\varepsilon^+)^2-(\varepsilon^-)^2)\varepsilon^+}
 +\frac{4\left(q^2+\Delta_1^2 \right)\Delta_3}{((\varepsilon^+)^2-(\varepsilon^-)^2)\varepsilon^-} \right).
\label{Delta3_gap_equation}
\end{eqnarray}
If one assumes $\Delta_3=0$ the first equation reduces to the gap equation
solved in \cite{Huang_propagator}.  The second equation does not vanish
exactly under this assumption.  If one takes $m=0$ and $\Delta_3=0$ the second equation is 
trivially satisfied and the first equation becomes the gap equation for 
massless quarks.

The range of integration for $q$ is not infinite 
since the NJL model is a four-fermion interaction model and must
have an UV cutoff, $\Lambda$, which is left arbitrary except for
the restriction that it be greater than $\mu$.

An approximate solution of the gap equation
for $m=0$ is given by\cite{Review2}:
\begin{equation}
\Delta_1^{(0)}\equiv
\Delta_1(m=0)\approx 2 \sqrt{\Lambda^2-\mu^2} \exp\left[{\frac{\Lambda^2-3\mu^2}{2\mu^2}}\right]
\exp\left[{-\frac{1}{2 G \,\mu^2}}\right].
\label{m=0solution}
\end{equation}

Expanding the integrands of (\ref{Delta1_gap_equation}) and (\ref{Delta3_gap_equation}) in 
$m$ one can obtain approximate solutions for the condensates: 
\begin{eqnarray}
\Delta_1(m)&\!\!\!\!\!\!\!\!\!\approx&\!\!\!\!\!\!\!\!\!\Delta_1^{(0)}\left[ 1 - m^2\left\{ \frac{1-G(\Lambda^2-\mu^2)}{4\, G \, \mu^4 }
-\frac{1}{2(\Lambda^2-\mu^2)}
+ A \frac{1-G(\Lambda^2+\mu^2)}{2\, G\, \mu^4}\right\}\right. \nonumber\\
&+& \frac{1}{2} m^4\left\{ \frac{1-G(\Lambda^2-\mu^2)}{4\, G \, \mu^4 }
-\frac{1}{2(\Lambda^2-\mu^2)}
+ A \frac{1-G(\Lambda^2+\mu^2)}{2\, G\, \mu^4}\right\}^2 \nonumber\\
&+& m^4 \left\{ \frac{3 (G(\Lambda^2-\mu^2)-1) }{16 G \mu^6 }-\frac{1}{4(\Lambda^2-\mu^2)^2}
+\frac{\log[\frac{\Lambda}{\mu}]}{8\mu^4}
\right\} \nonumber\\
&+& \left. A\,m^4 \left\{ \frac{(G(\Lambda^2+\mu^2)-1) }{4 G \mu^6 }+\frac{A(G \Lambda^2-1)}{2\,G\,\mu^6}
-\frac{\log[\frac{\Lambda}{\mu}]}{2\mu^4}
\right\}  \right],\\
\Delta_3(m)&\approx& A \;\frac{m}{\mu} \; \Delta_1(m),
\end{eqnarray}
where:
\begin{equation}
A=1- \frac{3 \Lambda^2-\mu^2}{3 \Lambda^2-7 \mu^2+3\mu^2 \log\left[\left(\Delta_1^{(0)}\right)^2/4(\Lambda^2-\mu^2)\right]}\;.
\end{equation} 
In order to verify the accuracy of the approximate solutions the exact gap
equations were also solved numerically using {\it Mathematica}$\,^\copyright$
for comparison.  These approximations are somewhat complicated due to
their generality and range of applicability so some general comments
are in order.

The solution for $\Delta_3$ is linear in $m/\mu$ for small values
of m with slope $A$ ranging from $A=1$ for $\Delta_1^{(0)}=0$ to 
$A= 0$ for  $\Delta_1^{(0)}=\frac{2}{e}\sqrt{\Lambda^2-\mu^2}$.

The solution for $\Delta_1(m)$ is an expansion in $\frac{m^2}{\mu^2}$
to order ${\cal O}(m^4/\mu^4)$.  Factors of $1/G \mu^2$ in some terms
are actually enhancement factors as can be seen by taking typical
values of $G=6 \,\mbox{GeV}^{-2}/\pi^2$ and $\mu=500$ MeV giving a
value of $1/G \mu^2\approx 6.5$.  Expanding only to ${\cal O}(m^2/\mu^2)$:
\begin{equation}
\Delta_1(m)\!=\!\Delta_1^{(0)}\!\left[ 1 - m^2\left\{ \frac{1-G(\Lambda^2-\mu^2)}{4\, G \, \mu^4 }
-\frac{1}{2(\Lambda^2-\mu^2)}
+ A \frac{1-G(\Lambda^2+\mu^2)}{2\, G\, \mu^4}\right\}\right]\!.
\end{equation}
it can be shown that the first and last terms dominate and are of the
same order of magnitude. This means that neglecting the effect of
$\Delta_3$ by setting $A=0$ will have a non-trivial effect on the
solution for $\Delta_1(m)$. The order ${\cal O}(m^4/\mu^4)$ terms are
less instructive and are only included in order that the approximation
is accurate for a wider range of $m$, $\mu$, $\Lambda$ and $G$.  

\begin{figure}[t]
\epsfysize=2.7in
\epsfbox[95 22 376 165]{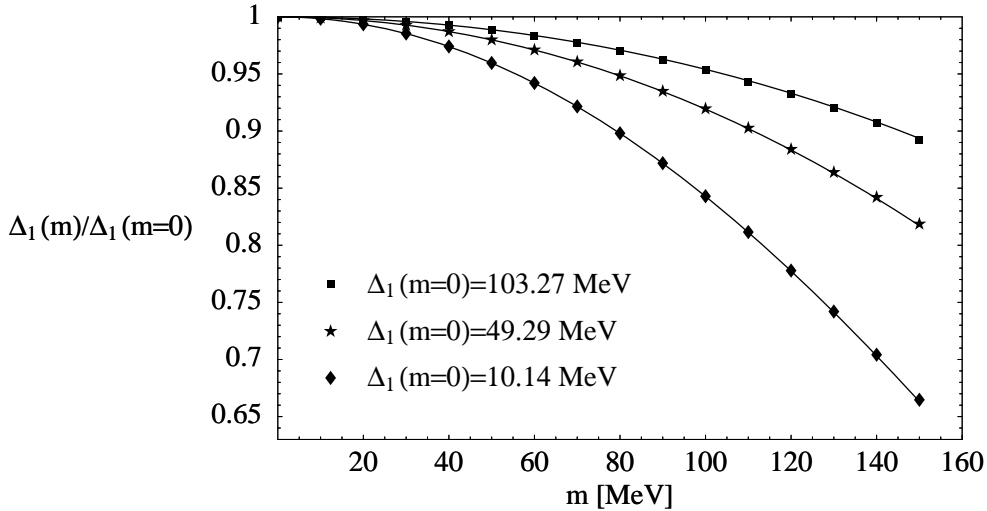}
\caption{$\Delta_1(m)$ for $\mu=500$ MeV and $\Lambda=2\mu$ in the scalar NJL 
model for (from bottom to top) $G \pi^2 [\mbox{GeV}^{-2}]=3.5,4.875,6$. 
Analytic approximations and exact numerical results are shown for comparison.} 
\label{scan_naive}
\end{figure}

The accuracy of this approximation for a range of four fermion couplings,
$G$, is illustrated in Figure (\ref{scan_naive}). Examining the
limiting case $G\rightarrow 0$ which corresponds to $\Delta_1^{(0)}
\rightarrow 0$, it can be seen that effect of the quark mass becomes
increasingly important.  Conversely, as $\Delta_1^{(0)}$ increases the
effect of the quark mass becomes less significant.  Results obtained
by varying $\Lambda$ or $\mu$ have exactly the same behavior and also
agree well with the exact numerical results.

\begin{figure}[hb]
\epsfysize=2in
\epsfbox[10 95 277 195]{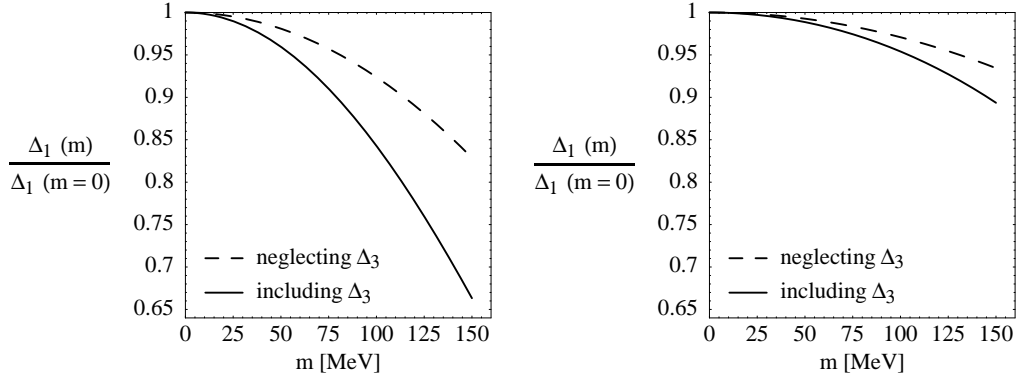}
\caption{$\Delta_1(m)$ for $\mu=500$ MeV and $\Lambda=2\mu$ in the 
scalar NJL model.  The graph on the left is for 
$G=3.5\;\mbox{GeV}^{-2}/\pi^2$ and $\Delta_1^{(0)}=10.14$ MeV.  The graph
on the right is for $G=6\; \mbox{GeV}^{-2}/\pi^2$ and $\Delta_1^{(0)}=103.27$ MeV.
Solutions are shown neglecting and including $\Delta_3$.}
\label{Delta1_naive}
\end{figure}

\begin{figure}[h]
\epsfysize=2.7in
\epsfbox[93 22 376 165]{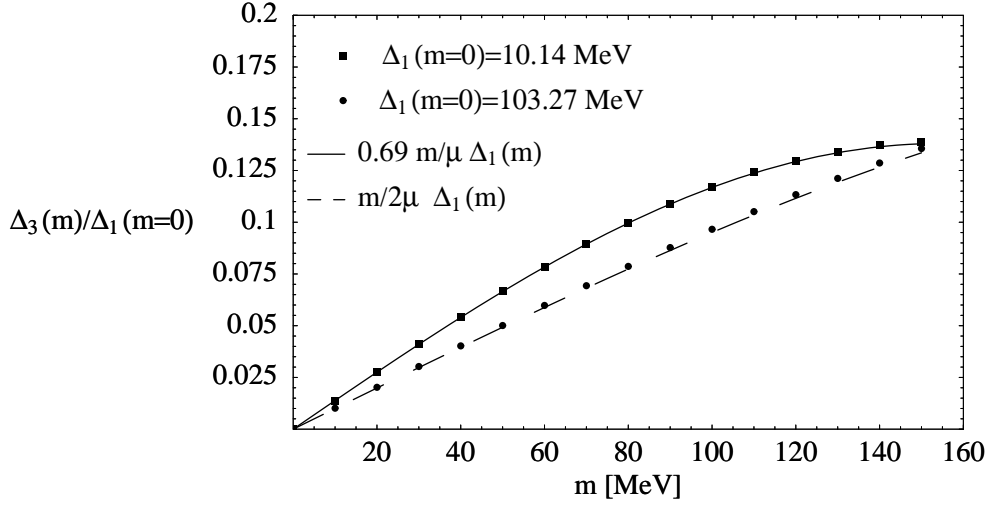}
\caption{$\Delta_3(m)$ for $\mu=500$ MeV and $\Lambda=2\mu$  in the 
scalar NJL model. Four fermion coupling constants 
$G=3.5\; \mbox{GeV}^{-2}/\pi^2$ and $G=6\;\mbox{GeV}^{-2}/\pi^2$ were used
and give solutions with $\Delta_1^{(0)}=10.14$ MeV and
$\Delta_1^{(0)}=103.27$ MeV.  Analytic approximations and exact
numerical results are shown for both cases.}
\label{Delta3_naive}
\end{figure}

The solutions for $G=3.5 \, \mbox{GeV}^{-2}/\pi^2$ and $G=6 \,
\mbox{GeV}^{-2}/\pi^2$ which correspond to $\Delta_1^{(0)}=10.14$ MeV and 
$\Delta_1^{(0)}=103.27$ MeV are shown in Figures
(\ref{Delta1_naive}) and (\ref{Delta3_naive}) with $\mu=500$ MeV.
Also shown in Figure (\ref{Delta1_naive}) are the solutions where
$\Delta_3$ is neglected for comparison.  The effect on $\Delta_1(m)$
of including $\Delta_3$ is an increase in the mass dependence compared
to the solution neglecting $\Delta_3$.  The effect is more pronounced
for smaller values of $\Delta_1^{(0)}$, but is still significant at 
$\Delta_1^{(0)}\approx 100$ MeV.  Notice that the relative size of
$\Delta_3$ with respect to $\Delta_1$ is very similar in both cases.

\begin{figure}[h]
\epsfysize=2.2in
\epsfbox[98 19 365 128]{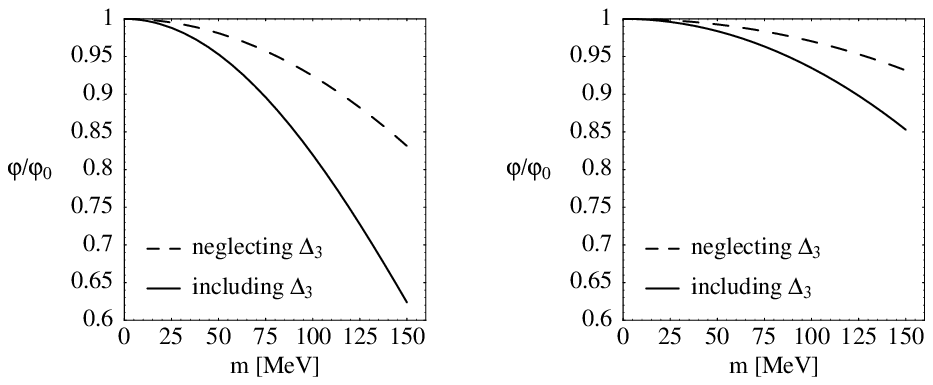}
\caption{Plot of $\varphi/\varphi_0$ for the scalar NJL model including and neglecting $\Delta_3$
for different values of $\Delta_1^{(0)}$.  The graph on the left is for
$G=3.5\; \mbox{GeV}^{-2}/\pi^2$ and $\Delta_1^{(0)}=10.14$ MeV.  The
graph on the right is for $G=6\;\mbox{GeV}^{-2}/\pi^2$ and
$\Delta_1^{(0)}=103.27$ MeV.  
}
\label{gapplot_naive}
\end{figure}

As discussed in the previous section, inclusion of $\Delta_3$ alters
the dispersion relations and the gap, $\varphi$. The gap between the
quasiparticle hole branch and quasiparticle branch in this case is not
simply $\varphi=2\Delta_1(m)$ but is given by (\ref{gap}).  The value
of the gap is plotted for the same choices of $G$ presented above in
Figure (\ref{gapplot_naive}) with and without $\Delta_3$.  The effect
on the gap is larger than the effect on $\Delta_1$ alone because
$\Delta_3$ affects $\varphi$ directly in (\ref{gap}) as well as
indirectly through $\Delta_1$.  The dispersion relation for the
quasiparticle holes is not appreciably altered.

\newpage

\newpage

\newpage
\newpage

\section{NJL Model motivated by Single Gluon Exchange}

The NJL interaction which has the structure of single gluon exchange
has the interaction vertices:
\begin{eqnarray}
\Gamma^A_\mu= \gamma_\mu \frac{\lambda^A_c}{2},~~~~& ~~~~
{\bar \Gamma}^A_\mu= -\gamma_\mu (\frac{\lambda^A_c}{2})^T,
\end{eqnarray}
and:
\begin{equation}
g^2 D^{\mu\nu}_{AB} \rightarrow 3\pi^2 G g^{\mu\nu} \delta_{AB},
\end{equation}
giving the gap equation:
\begin{equation}
\Delta(k)= -2\; i\; G \int \frac{d^4q}{(2 \pi)^4}\; \gamma_\nu \;  G^-_0(q) \Delta(q) G^+(q)\; \gamma^\nu,
\label{GapEquation3}
\end{equation}
using the equations:
\begin{equation}
(\lambda^A)^T_{ij}
 \lambda^B_{jk} \lambda^A_{kl}
= -\frac{8}{3} \lambda^B_{il} ~~~~\mbox{assuming $\lambda^B$ is antisymmetric.}
\end{equation}
Multiplying each side of equation (\ref{GapEquation3}) by $\gamma_5$ and $\gamma_5\gamma_0$,
tracing over the Dirac indices and using the cyclicity of the trace and the relations:
\begin{equation}
\gamma_\mu \gamma^5 \gamma^\mu= -4 \gamma^5,
\end{equation}
\begin{equation}
\gamma_\mu \gamma^5 \gamma^0 \gamma^\mu= - \gamma^5 \gamma_\mu \gamma^0
\gamma^\mu= 2\gamma^5 \gamma^0,
\end{equation}
and then applying all the other machinery of the last section,
the coupled gap equations are obtained:
\begin{equation}
\Delta_1= G\int dq \; q^2\;\left(\frac{\Delta_1}{\varepsilon^+} + \frac{\Delta_1}{\varepsilon^-}+\frac{4(m\; \mu+\Delta_1\Delta_3)\Delta_3}
{((\varepsilon^+)^2-(\varepsilon^-)^2)\varepsilon^+} -\frac{4(m\; \mu+\Delta_1\Delta_3)\Delta_3}
{((\varepsilon^+)^2-(\varepsilon^-)^2)\varepsilon^-} \right),
\end{equation}
\begin{eqnarray}
\Delta_3\!\!&=& \!\!-\frac{G} {2}\int dq \; q^2\;\left(-\frac{\Delta_3}{\varepsilon^+} - \frac{\Delta_3}{\varepsilon^-}-\frac{4\;m\; \mu\;\Delta_1}
{((\varepsilon^+)^2-(\varepsilon^-)^2)\varepsilon^+} +\frac{4\;m\; \mu\;\Delta_1}
{((\varepsilon^+)^2-(\varepsilon^-)^2)\varepsilon^-} \right.  \nonumber \\
&-&\left. \frac{4\left(q^2+\Delta_1^2 \right)\Delta_3}{((\varepsilon^+)^2-(\varepsilon^-)^2)\varepsilon^+}
 +\frac{4\left(q^2+\Delta_1^2 \right)\Delta_3}{((\varepsilon^+)^2-(\varepsilon^-)^2)\varepsilon^-} \right).
\end{eqnarray}
Notice that they are almost the same as the equations in the last section except for
an overall factor of $-\frac{1}{2}$ in the second equation.  Therefore the only change to
the approximate solutions of the last section is:
\begin{equation}
A=-\frac{1}{2}+ \frac{3 \Lambda^2-4\mu^2}{3 \Lambda^2-10 \mu^2+3\mu^2 \log\left[\left(\Delta_1^{(0)}\right)^2/4(\Lambda^2-\mu^2)\right]}\;.
\end{equation} 

\begin{figure}[h]
\epsfysize=2.7in
\epsfbox[95 22 376 165]{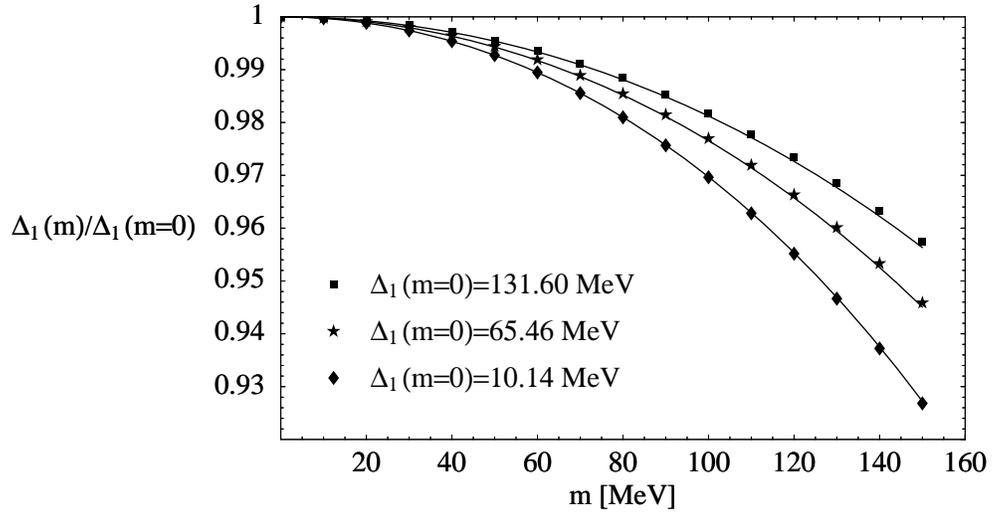}
\caption{$\Delta_1(m)$ for $\mu=500$ MeV and $\Lambda=2\mu$ in the 
NJL model with the color structure of single gluon exchange for (from bottom to top)  
$G \pi^2 [\mbox{GeV}^{-2}]=3.5,5.25,6.5$. Analytic approximations and exact
numerical results are shown for comparison.}
\label{scan_SGE}
\end{figure}

\begin{figure}[h]
\epsfysize=2.7in
\epsfbox[95 22 376 165]{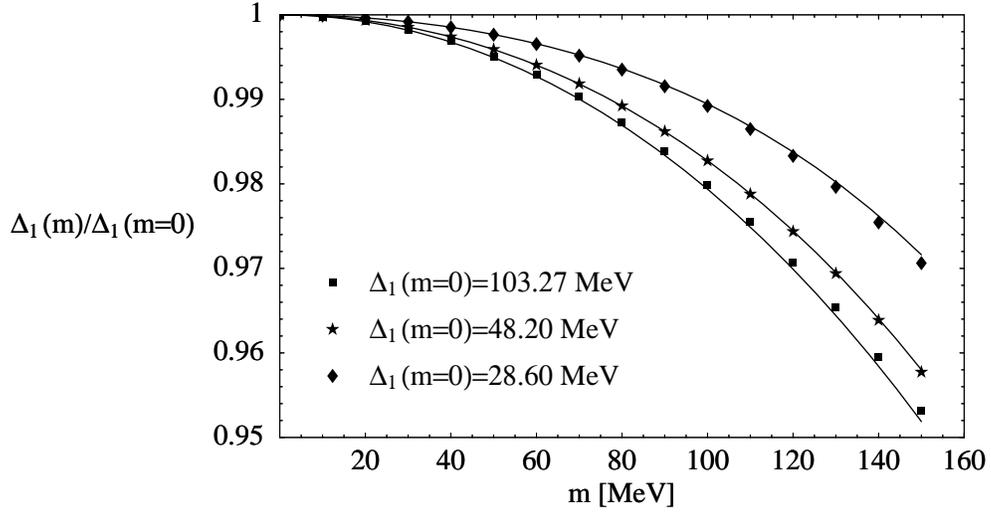}
\caption{$\Delta_1(m)$ for $\mu=500$ MeV and $G=6\;\mbox{GeV}^{-2}/\pi^2$ in the 
NJL model with the color structure of single gluon exchange for (from bottom to top) $\Lambda=2 \mu, 1.7 \mu,1.5 \mu$. Analytic approximations and exact
numerical results are shown for comparison.}
\label{Lambda_scan_SGE}
\end{figure}

For clarity the solution
to order ${\cal O}(m^2/\mu^2)$ is repeated here:
\begin{equation}
\Delta_1(m)\!=\!\Delta_1^{(0)}\!\left[ 1 - \!m^2\!\left\{\! \frac{1-G(\Lambda^2-\mu^2)}{4\, G \, \mu^4 }
-\frac{1}{2(\Lambda^2-\mu^2)}
+ A \frac{1-G(\Lambda^2+\mu^2)}{2\, G\, \mu^4}\right\} \right]\!,
\label{approx}
\end{equation}
\begin{equation}
\Delta_3(m)= A \;\frac{m}{\mu} \; \Delta_1(m).
\label{Delta3approx}
\end{equation} 
The accuracy of this approximation for a range of these
parameters is illustrated in Figures (\ref{scan_SGE})-(\ref{Lambda_scan_SGE})
by comparison with exact numerical solutions.

The main difference from the previous section is in the sign of
A. $\Delta_3(m)$ is again linear in $m/\mu$ for small m but with a
slope of $A\approx -\frac{1}{2}$ for small values of $\Delta_1^{(0)}$
and approaching $0$ as $\Delta_1^{(0)}$ increases.

The first and third terms in the ${\cal O}(m^2/\mu^2)$ expansion of
$\Delta_1(m)$ are of the same order of magnitude but in this case
their signs differ.  The resulting effect of the inclusion of
$\Delta_3$ is to reduce the relative effect of the mass dependence.

The comments on the mass dependence as a function of the parameters
$\mu$ and $G$ are unchanged from the previous section.  For small
values of $\Delta_1^{(0)}$ the effect is important and is less
significant for increasing $\Delta_1^{(0)}$.  This effect can be seen
in Figure (\ref{scan_SGE}).

In the weak coupling limit, $G \rightarrow 0$, equations (\ref{approx}) and 
(\ref{Delta3approx}) are very similar to equations (37) and (34) of
\cite{MassTerms}.

The mass dependence of $\Delta_1$ as a function of $\Lambda$ is
different in this model than in the last section.  In this case the
last two terms of equation (\ref{approx}) decrease as $\Lambda$
approaches $\mu$ and although the first term still dominates, the
relative mass dependence decreases with decreasing $\Lambda$.  The
effect is that the mass dependence is less significant for smaller
$\Delta_1^{(0)}$ as can be seen in Figure (\ref{Lambda_scan_SGE}).

\begin{figure}[h]
\epsfysize=2in
\epsfbox[10 95 277 195]{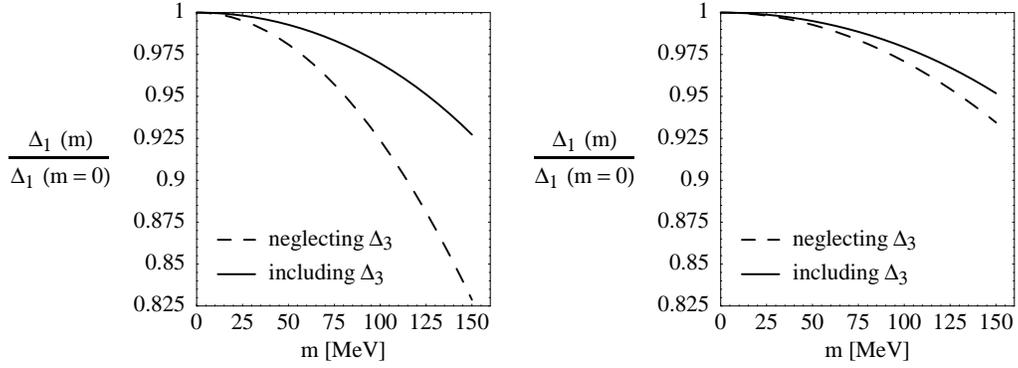}
\caption{$\Delta_1(m)$ for $\mu=500$ MeV and $\Lambda=2\mu$ in the 
NJL model with the color structure of single gluon exchange. 
The graph on the left is for 
$G=3.5\;\mbox{GeV}^{-2}/\pi^2$ and $\Delta_1^{(0)}=10.14$ MeV.  The graph
on the right is for $G=6\; \mbox{GeV}^{-2}/\pi^2$ and $\Delta_1^{(0)}=10.14$ MeV.
Solutions are shown neglecting and including $\Delta_3$.}
\label{Delta1_SGE}
\end{figure}

\begin{figure}[h]
\epsfysize=2.7in
\epsfbox[93 22 376 165]{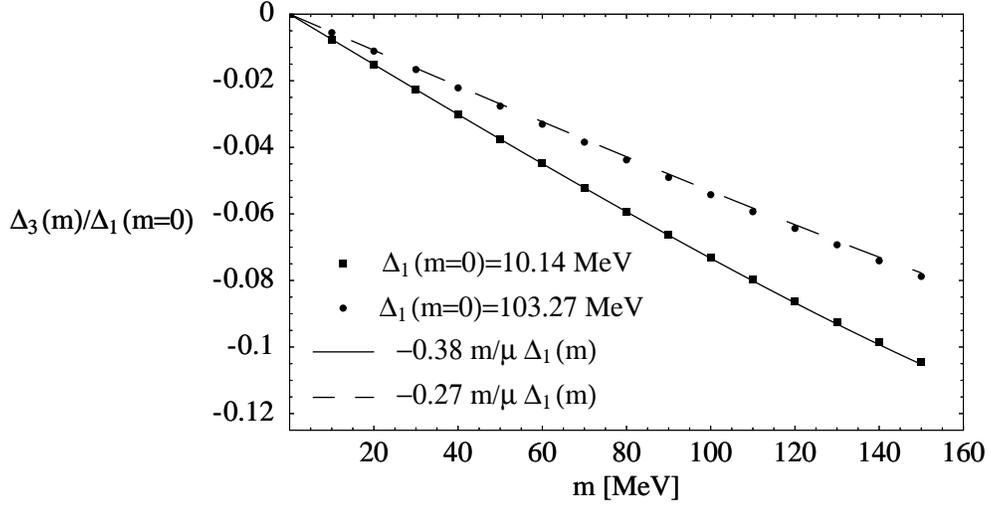}
\caption{$\Delta_3(m)$ for $\mu=500$ MeV and $\Lambda=2\mu$ in the 
NJL model with the color structure of single gluon exchange. Four fermion coupling constants 
$G=3.5\;\mbox{GeV}^{-2}/\pi^2$ and $G=6\; \mbox{GeV}^{-2}/\pi^2$ were used
and give solutions with $\Delta_1^{(0)}=10.14$ MeV and
$\Delta_1^{(0)}=103.27$ MeV.  Analytic approximations and exact
numerical results are shown for both cases.}
\label{Delta3_SGE}
\end{figure}

The solutions for $G=3.5 \,\mbox{GeV}^{-2}/\pi^2$ and $G=6 \,
\mbox{GeV}^{-2}/\pi^2$ which correspond to $\Delta_1(m=0)=10.14$
MeV and $\Delta_1(m=0)=103.27$ MeV are shown in Figures
(\ref{Delta1_SGE}) and (\ref{Delta3_SGE}) with $\mu=500$ MeV.  The
solutions where $\Delta_3$ is neglected are also shown in Figure
(\ref{Delta1_SGE}) for comparison.  The effect on $\Delta_1(m)$ of
including $\Delta_3$ is a reduction in the relative effect compared to
the solution neglecting $\Delta_3$.    The effect is more pronounced
for smaller values of $\Delta_1^{(0)}$, but is still nonneglegible at 
$\Delta_1^{(0)}\approx 100$ MeV.  Notice that the relative size of
$\Delta_3$ with respect to $\Delta_1$ is very similar in both cases.

\begin{figure}[h]
\epsfysize=2.2in
\epsfbox[98 19 365 128]{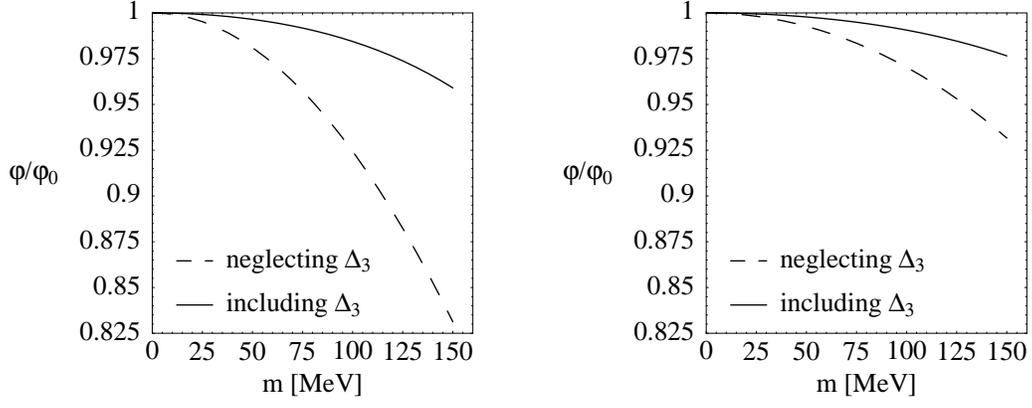}
\caption{Plot of $\varphi/\varphi_0$ for the NJL model with color structure of single
gluon exchange, including and neglecting $\Delta_3$ for different
values of $\Delta_1^{(0)}$.  The graph on the left is for $G=3.5\;
\mbox{GeV}^{-2}/\pi^2$ and $\Delta_1^{(0)}=10.14$ MeV.  The graph on
the right is for $G=6\;\mbox{GeV}^{-2}/\pi^2$ and
$\Delta_1^{(0)}=103.27$ MeV.  
}
\label{gapplot_SGE}
\end{figure}

Inclusion of $\Delta_3$ alters the dispersion relations and the
gap. The gap between the quasiparticle hole branch and quasiparticle
branch is not simply $\varphi=2\Delta_1$, but is again given by
(\ref{gap}).  The gap depends on $\Delta_3$ both explicitly and
implicitly through its effect on $\Delta_1$.  The value of the gap is
plotted for the same choices of $G$ presented above in Figure
(\ref{gapplot_SGE}) with and without $\Delta_3$.  The effect on the
gap is larger than the effect on $\Delta_1$ alone.  The zeroth order
approximation, $\Delta_1(m=0)$, is a better approximation to the gap
than the solution where $\Delta_3$ is neglected.  The dispersion
relation for the quasiparticle holes is not appreciably altered.

Extending this analysis out to quark mass of $m=150$ MeV is partly for
illustrative purposes but is also valuable for two other reasons.

First the analysis of \cite{Huang_propagator} shows that in a coupled
analysis of the superconducting (diquark) condensate and the axial
condensate, constituent quark masses of the order of $100$ MeV for the
light quarks are possible at $\mu\approx 400$ MeV.  For constituent
quark masses of this order the results of our analysis show that the
presence of a new condensate, $\Delta_3$ which they neglected in their
ansatz will be relevant.  Their ansatz is a reasonable first
approximation as the new condensate and its effects on $\Delta_1$ are
not large. However, the effect on the gap $\varphi$ itself is
significant even for gaps of the order of $100$ MeV.  The full
analysis requires the more general ansatz and the use of the general
quasiparticle propagator.  The effects of chiral symmetry breaking
have not been taken into account in this paper as in
\cite{Huang_propagator,Buballa_Hosek_Oertel} but
such an extension would not be difficult.  The authors of
\cite{Buballa_Hosek_Oertel} also found significant values for
the constituent quark masses for $\mu\approx 400$ MeV.

Second the methods used in this analysis can be extended to the physical
case where the strange quark mass is of the order of $150$ MeV.  
It is instructive therefore to carry out this analysis as a precursor
to the physical case.

\begin{figure}[ht]
\epsfysize=2.2in
\epsfbox[98 19 365 128]{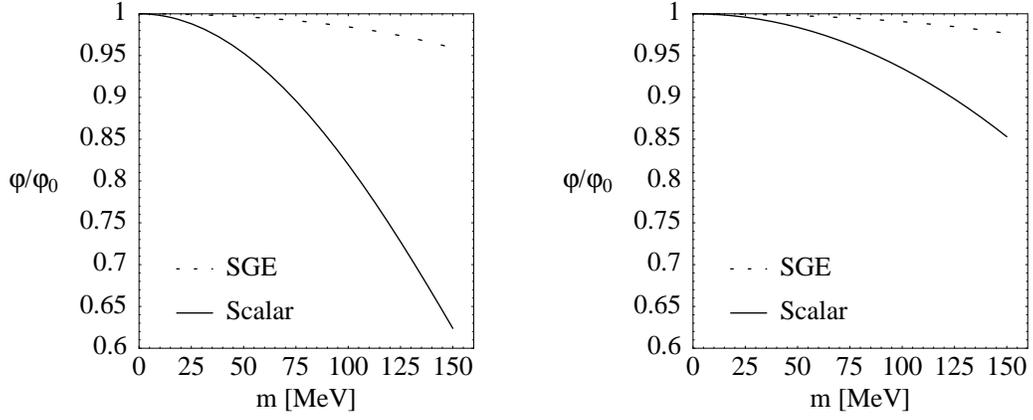}
\caption{Comparison of the full gap calculations in the scalar NJL model
and the single gluon exchange motivated NJL model.  The graph on the left is for 
$G=3.5\;\mbox{GeV}^{-2}/\pi^2$ and $\Delta_1^{(0)}=10.14$ MeV.  The graph on
the right is for $G=6\;\mbox{GeV}^{-2}/\pi^2$ and
$\Delta_1^{(0)}=103.27$ MeV.}
\label{gapplot_both}
\end{figure}

The two NJL models analyzed in this paper lead to 
quite different results when the full calculation is
performed.  In Figure (\ref{gapplot_both}) it is shown
that the full calculation depends significantly on the
type of NJL model used.  For lower values of $\Delta_1^{(0)}$
the effect on the gap is almost a factor of 2 and is still significant
even for $\Delta_1^{(0)}$ of the order of 100 MeV.
The second NJL model analyzed is perhaps more physically
motivated than the second but there is 

\newpage

\newpage

\newpage

\newpage

\newpage

\newpage
\newpage

\newpage

\noindent is no reason to restrict
the analysis to interactions of this form.  The most general
NJL model would involve both types of interactions.  This
analysis shows that there could be a significant dependence on 
the specific NJL model used.

\section{Conclusion}

In this paper the general quasiparticle propagator for the
case of equally massive quarks is derived.  The quasiparticle propagator
is then specialized to the propagator in an NJL model.  It is shown
that there is exactly one new condensate, $\Delta_3$, in this model.

The quasiparticle propagator was used to solve the gap equations for
the CFL condensate, $\Delta_1$, and the new condensate, $\Delta_3$ in
two NJL models as a function of the quark mass. Approximate analytic
solutions were obtained in both cases valid over a range of values of
the parameters $m$, $\Lambda$, $\mu$ and $G$ in both models.  The
accuracy of these approximations was tested by numerically solving the
exact gap equations.  This represents a significant advance as the
analytic approximations render numerical solution of the gap equations
unnecessary for a significant range of parameter space.

Results for the condensates, $\Delta_1$ and $\Delta_3$, and the gap,
$\varphi$, in the scalar NJL model were presented as a function of
the quark mass.  The full solution for $\Delta_1$ and $\varphi$ were compared
to solutions obtained neglecting $\Delta_3$.  This was
done for different values of the four fermion coupling constant.  The
results show that the complete solution differs from the approximate
solution especially at large quark mass and smaller $\Delta_1$. 

Results for the condensates, $\Delta_1$ and $\Delta_3$, and the gap,
$\varphi$, in the NJL model with the structure of single gluon
exchange were presented as a function of the quark mass. The full
solution for $\Delta_1$ and $\varphi$ were compared to solutions
obtained neglecting $\Delta_3$.  This was again done for different
values of the four fermion coupling constant.  The results show that
the complete solution differs from the approximate solution especially
at large quark mass and smaller $\Delta_1$.

The solutions obtained in this paper are relevant to the work of
\cite{Huang_propagator} where $\Delta_3$ was neglected as a first approximation.
The approach of this paper could be used to generalize their analysis
to the complete case.

The authors of \cite{Buballa_Hosek_Oertel} showed that $\Delta_3$
could be quite large in an NJL model with a Lorentz non-invariant
interaction and have a significant effect on the gap.  In the analysis
presented here it is shown that even for small values of $\Delta_3$,
neglecting this condensate can have a non-trivial effect on the gap.

Comparison of the results for the gap in the two different
NJL models analyzed in this paper show that the gap is
strongly dependent on the type of NJL model used.

Solving the color superconducting gap equations for the case of equal
mass quarks is also valuable as a precursor to analyzing the physical
case where the up and down quarks are essentially massless and the
strange quark is massive.  The methods used in this research combined
with the methods of \cite{Fugleberg} can be combined to analyze the
physical case
which is the ultimate goal of this line of research.

\vspace{0.2in}
{\bf Acknowledgments}
\vspace{0.1in}

I would like to thank D. Rischke, R. Pisarski, T. Sch\"{a}fer and
J. Lenaghan for many valuable discussions and comments.  This research
was funded by a Natural Science and Engineering Research Council
(NSERC) of Canada Post Doctoral Fellowship.  This work was also
supported in part by DOE grant DE-AC02-98CH10886.

\appendix

\section{Nambu-Gorkov Formalism}

\setcounter{equation}{0}
\renewcommand{\theequation}{\thesection \arabic{equation}}
In the standard Nambu-Gorkov formalism one deals with the eight component
spinors:
\begin{eqnarray}
\Psi\equiv \left( 
\begin{array}{c} \psi \\ \psi_C 
\end{array} \right),
&
\bar{\Psi}\equiv \left( 
 \psi ~~ \psi_C 
 \right),
\end{eqnarray}
where $C$ is the charge conjugation operator.
The action for interacting fermions in the mean field approximation can be 
written concisely as:
\begin{equation}
I[\Psi,\bar{\Psi}]=\frac{1}{2} \int \bar{\Psi}\; {\cal S}^{-1}\; \Psi,
\label{action_appendix}
\end{equation}
where:
\begin{equation}
{\cal S}^{-1} = \left(
\begin{array}{cc}
[G^+_0]^{-1} & \Delta^- \\
\Delta^+ & [G^-_0]^{-1}
\end{array}
\right).
\end{equation}
$G^\pm_0$ are the free propagators of particles and charge-conjugate particles, $\Delta^+$ is a
matrix of diquark condensates and $\Delta^-\equiv \gamma_0 \Delta^+ \gamma_0$.

The full Nambu-Gorkov propagator is:
\begin{equation}
{\cal S} = \left(
\begin{array}{cc}
G^+ & -G^+_0\Delta^- G^-\\
-G^-_0\Delta^+ G^+ & G^-
\end{array}
\right),
\label{def_full_propagator}
\end{equation}
where:
\begin{equation}
G^\pm\equiv\left\{ [G^\pm_0]^{-1}-\Delta^\mp G^\mp_0 \Delta^\pm\right\}^{-1}.
\label{quasiparticle_propagator_Appendix}
\end{equation}
are the full quasiparticle and charge-conjugate quasiparticle propagators.
The gap equation is a consistency equation\cite{RD_superfluid}:
\begin{equation}
\Delta^+(k)= - i g^2 \int \frac{d^4q}{(2 \pi)^4} \sum_{A,B=1..N_c^2-1} {\bar \Gamma}^A_\mu D^{\mu\nu}_{AB}(k-q)
 G^-_0(q) \Delta^+(q) G^+(q) \Gamma^B_\nu.
\label{GapEquation1_Appendix}
\end{equation}
which must be satisfied in order that the effects of interaction at the
mean field level are consistently incorporated in the action (\ref{action_appendix}) 
in terms of the diquark condensate matrix\footnote{$\Delta^+$ is also called the
gap matrix since it elements determine the gaps in the quasiparticle spectrum.}, $\Delta^+$.
$D^{\mu\nu}_{AB}(k-q)$ is the gluon propagator,
$\Gamma^B_\nu$ is the interaction vertex, and:
\begin{equation}
{\bar \Gamma}= C  \Gamma^T  C^{-1}.
\end{equation}

\setcounter{equation}{0}

\section{Dirac Basis}

The Bailin and Love motivated decomposition given in \cite{RD_superfluid} is:
\begin{equation}
\Delta\!= \!\Delta_1 \gamma_5 + \Delta_2 \gamma\cdot\hat{k}\gamma_0\gamma_5
+ \Delta_3 \gamma_0\gamma_5+ \!\Delta_4 + \Delta_5 \gamma\cdot\hat{k}\gamma_0
+ \Delta_6 \gamma\cdot\hat{k}+ \Delta_7 \gamma\cdot\hat{k}\gamma_5 
+\Delta_8 \gamma_0,
\end{equation}
where:
\begin{equation}
\Delta_1 \gamma_5 + \Delta_2 \gamma\cdot\hat{k}\gamma_0\gamma_5,
\end{equation}
represents condensation of fermions with the same chirality in the
even parity channel,
\begin{equation}
\Delta_4 + \Delta_5 \gamma\cdot\hat{k}\gamma_0,
\end{equation}
represents condensation of fermions with the same chirality in the
odd parity channel.
\begin{equation}
\Delta_3 \gamma_0\gamma_5+ \Delta_7 \gamma\cdot\hat{k}\gamma_5,
\end{equation}
represents condensation of fermions with opposite chirality in the
even parity channel,
\begin{equation}
\Delta_6 \gamma\cdot\hat{k}+\Delta_8 \gamma_0,
\end{equation}
represents condensation of fermions with opposite chirality in the
odd parity channel.


\begin{thebibliography}{99}
\bibitem{Barrois} B.C. Barrois, Nucl.Phys. {\bf B129} (1977) 390.
\bibitem{BailinLove} D. Bailin and A. Love, Phys. Rep. {\bf 107} (1984) 325.
\bibitem{ARW_RSSV} M. Alford, K. Rajagopal and F. Wilczek Phys.Lett. {\bf B422} (1998) 247 [hep-ph/9711395]; R. Rapp, T. Schaefer, E. Shuryak and M. Velkovsky, Phys.Rev.Lett. {\bf 81} (1998) 53 [hep-ph/9711396].
\bibitem{Review1} T. Sch\"{a}fer and E. Shuryak, Lect.Notes Phys. 578 (2001) 203 [nucl-th/0010049].
\bibitem{Review2} K. Rajagopal and F. Wilczek, {\it The Condensed Matter Physics of QCD}, Chapter 35 in 
the Festschrift in honor of B. L. Ioffe, "At the Frontier of Particle Physics / Handbook of QCD", M. Shifman, ed., (World Scientific) [hep-ph/0011333].
\bibitem{CCSCCompact} M. Alford, J. Bowers and K. Rajagopal, J.Phys.{\bf G27} (2001) 541 [hep-ph/0009357].
\bibitem{Carter_Blinking} G.W. Carter, {\it Color Superconductivity and Blinking Proto-Neutron Stars}, hep-ph/0111353.
\bibitem{Prakash} M. Prakash, Nucl.Phys. A698 (2002) 440, [hep-ph/0105158].
\bibitem{Pons_etal} J. Pons, A. Steiner, M. Prakash, J. Lattimer, Phys.Rev.Lett. 86 (2001) 5223,[astro-ph/0102015].
G.W. Carter and S. Reddy, Phys.Rev. {\bf D62} (2000) 103002 [hep-ph/0005228].
\bibitem{CFL} M. Alford, K. Rajagopal and F. Wilczek, Nucl.Phys. {\bf B537} (1999) 443 [hep-ph/9804403].
\bibitem{CFL_accepted} T. Sch\"{a}fer and F. Wilczek, Phys.Rev.Lett. {\bf 82} (1999) 3956 [hep-ph/9811473]; \\
R. Rapp, T. Sch\"{a}fer, E.V. Shuryak and M. Velkovsky, Annals Phys. {\bf 280} (2000) 35 [hep-ph/9904353];\\
 T. Sch\"{a}fer, Nucl.Phys. {\bf B575} (2000) 269 [hep-ph/9909574]; \\
I. Shovkovy and  L. Wijewardhana, Phys.Lett. {\bf B470} (1999) 189 [hep-ph/9910225]; \\ 
N. Evans, J. Hormuzdiar, S. Hsu and M. Schwetz, Nucl.Phys. {\bf B581} (2000) 391 [hep-ph/9910313].
\bibitem{Crystalline} M. Alford, J. Bowers and K. Rajagopal, Phys.Rev. {\bf D63}
 (2001) 074016 [hep-ph/0008208]; K. Rajagopal, Nucl.Phys. {\bf A702} (2002) 25 [hep-ph/0109135].
\bibitem{Stress} P.F. Bedaque and T. Sch\"{a}fer, Nucl.Phys. {\bf A697} (2002) 802 [hep-ph/0105150]; \\
P.F. Bedaque, Phys.Lett. {\bf B524} (2002) 137 [nucl-th/0110049].
\bibitem{Kaplan_Reddy} D. Kaplan and S. Reddy, Phys.Rev. {\bf D65} (2002) 054042 [hep-ph/0107265].
\bibitem{Fugleberg} T. Fugleberg, {\it Color Superconductivity with Non-Degenerate Quarks}, hep-ph/0112162.
\bibitem{Unlocking_ABR} M. Alford, J. Berges and K. Rajagopal, Nucl.Phys.
{\bf B558} (1999) 219 [hep-ph/9903502].
\bibitem{SW_Description} T. Sch\"{a}fer and F. Wilczek, Phys.Rev. {\bf D60} 
(1999) 074014 [hep-ph/9903503].
\bibitem{Enforced} K. Rajagopal and F. Wilczek Phys.Rev.Lett. {\bf 86}
 (2001) 3492 [hep-ph/0012039].
\bibitem{Diagrammatic}
J. Bowers, J. Kundu, K. Rajagopal and E. Shuster, Phys.Rev. {\bf D64} (2001) 014024 [hep-ph/0101067].
\bibitem{Buballa_Oertel} M. Buballa and M. Oertel, Nucl.Phys. {\bf A703} (2002) 770 [hep-ph/0109095].
\bibitem{Buballa_Hosek_Oertel} M. Buballa, J. Hosek and M. Oertel, Phys.Rev. D65 (2002) 014018 [hep-ph/0105079].
\bibitem{Huang_propagator} M. Huang, P. Zhuang and W. Chao, Phys.Rev. {\bf D65} (2002) 076012 [hep-ph/0112124].
\bibitem{Mass_induced} J. Kundu and K. Rajagopal, Phys.Rev. {\bf D65} (2002) 094022 [hep-ph/0112206].
\bibitem{MassTerms}  T. Sch\"{a}fer, Phys.Rev. {\bf D65} (2002) 074006 [hep-ph/0109052].
\bibitem{Opening} A. Leibovich, K. Rajagopal and E. Shuster, Phys.Rev. 
{\bf D64} (2001) 094005 [hep-ph/0104073].
\bibitem{Harada}  M. Harada and S. Takagi, Prog.Theor.Phys. {\bf 107} (2002) 561 [hep-ph/0108173].
\bibitem{RD_superfluid} R. Pisarski and D. Rischke, Phys.Rev. {\bf D60} (1999) 
094013 [nucl-th/9903023].
\end{thebibliography}
\end{document}